\documentclass[fleqn,usenatbib]{mnras}
\usepackage{newtxtext,newtxmath}
\usepackage[T1]{fontenc}
\usepackage{ae,aecompl}
\usepackage{multirow}
\usepackage{xcolor}
\usepackage{graphicx}	% Including figure files
\usepackage{amsmath}	% Advanced maths commands
\usepackage{amssymb}	% Extra maths symbols
\newcommand{\bz}{$\langle B_z \rangle$}

\title[Wideband characteristics of ECME from HD\,133880]{Unraveling the complex magnetosphere of the B star HD\,133880 via wideband observation of coherent radio emission}

\author[B. Das et al.]{
Barnali Das,$^{1}$\thanks{E-mail: barnali@ncra.tifr.res.in}
Poonam Chandra,$^{1}$
and Gregg A. Wade$^{2}$
\\
$^{1}$National Centre for Radio Astrophysics, Tata Institute of Fundamental Research,  Pune University Campus, Pune-411007, India\\
$^{2}$Department of Physics and Space Science, Royal Military College of Canada, PO Box 17000, Station Forces, Kingston, ON K7K 7B4, Canada
}

\date{Accepted XXX. Received YYY; in original form ZZZ}

\pubyear{2020}

\begin{document}
\label{firstpage}
\pagerange{\pageref{firstpage}--\pageref{lastpage}}
\maketitle

% Abstract of the paper
\begin{abstract}
HD\,133880 is one of the six hot magnetic stars known to produce coherent pulsed radio emission by the process of electron cyclotron maser emission (ECME). In this paper, we present observations of ECME from this star over a wide frequency range, covering nearly 300--4000 MHz with the Giant Metrewave Radio Telescope (GMRT) and the Karl G. Jansky Very Large Array (VLA). This study, which is the first of its kind, has led to the discovery of several interesting characteristics of the phenomenon and also of the host star. 
We find that the observable properties of ECME pulses, e.g. the time lag between right and left circularly polarized pulses, the amplitudes of the pulses, and their upper cut-off frequencies appear to be dependent on the stellar orientation with respect to the line of sight. We suggest that all these phenomena, which are beyond the ideal picture, can be attributed to a highly azimuthally asymmetric matter distribution in the magnetosphere about the magnetic field axis, which is a consequence of both the high obliquity (the angle between rotation axis and the magnetic field axis) of the star and the deviation of the stellar magnetic field from a dipolar topology.
\end{abstract}

% Select between one and six entries from the list of approved keywords.
% Don't make up new ones.

%%%%%%%%%%%%%%%%%%%%%%%%%%%%%%%%%%%%%%%%%%%%%%%%%%

%%%%%%%%%%%%%%%%% BODY OF PAPER %%%%%%%%%%%%%%%%%%

\section{Introduction}
Kilo-gauss strength, ordered magnetic fields of fossil origin are found in around 10\% of the OBA stars \citep[e.g.][]{grunhut2017}. These magnetic fields have relatively simple topologies and are approximately dipolar in most cases. Interaction of the magnetic field with the charged particles in the radiatively driven stellar winds, caused by mass loss from the outer layers of the star, leads to emission in the optical, UV, X-ray and also at the radio wavelengths \citep[e.g.][]{petit2013,trigilio2004}. For the latter, the emission mechanism is gyrosynchrotron in most cases. However a small number of the magnetic B stars are found to produce coherent radio emission by the process of electron cyclotron maser emission \citep[ECME;][]{trigilio2000,chandra2015,das2018,lenc2018,das2019a,das2019b,leto2019,leto2020}. 
%The characteristics of this emission which make it stand out from the smooth gyrosynchrotron continuum is that they have very high ($\sim100\%$) circular polarization and are highly directed because of which they are seen as pulses. 
The characteristics of this emission which make it stand out from the smooth gyrosynchrotron continuum are its very high ($\sim100\%$) circular polarization and its high directivity. As a consequence, ECME is expressed observationally as short, intense, highly circularly polarized pulses.
The pulses are periodic with a periodicity equal to the stellar rotation period. The rotational phases of arrival of the pulses are always close to the nulls of the stellar longitudinal magnetic field \bz, meaning that the radiation is directed roughly perpendicular to the magnetic field axis. The frequency of emission is proportional to the local electron cyclotron frequency which is a function of distance from the stellar surface. As a result, higher frequencies originate closer to the star and lower frequencies originate farther away from the star.
The radiation is believed to be generated in auroral rings near the magnetic poles \citep{trigilio2011}. For ECME in the extra-ordinary (X-) mode, right circularly polarized (RCP, according to the IEEE convention) radiation is produced near the star's north magnetic pole and left circularly polarized (LCP, according to the IEEE convention) radiation is produced near the south magnetic pole. On the other hand, if the dominant magneto-ionic mode is ordinary (O-), LCP radiation is produced near the star's north magnetic pole and RCP radiation is produced near the south magnetic pole. As a result, the pulse arrival sequence for the two circular polarizations are different for the two magneto-ionic modes. For example, when the north magnetic pole is receding and the south magnetic pole is approaching, we will see RCP followed by LCP in case of X-mode emission, and LCP followed by RCP in case of O-mode emission \citep[e.g.][]{leto2019}.% \citep[illustrated in Figure 1 of ][]{das2019a}.

The very first hot magnetic star from which ECME was discovered is CU\,Vir \citep{trigilio2000}. More than a decade later, the second such star, HD\,133880, was discovered. It was first speculated to be a host of this phenomenon by \citet{chandra2015} when they observed strong enhancements in flux density at 610 MHz and 1.4 GHz at certain rotational phases with the Giant Metrewave Radio Telescope (GMRT). This, however, could not be firmly established as they had a sparse rotational phase coverage and no polarization information. Eventually, ECME at 610 MHz was confirmed by \citet{das2018} when they observed the star for one full rotation cycle and found strong enhancement in flux density in RCP near one of the magnetic nulls. 
%Near the other magnetic null, data could not be taken due to technical issues. Note that \citet{das2018} reported the absence of LCP for HD\,133880 near the null covered by their data.

In this paper, we present wideband observations of HD\,133880 near its magnetic nulls. We used two radio telescopes: the upgraded GMRT (uGMRT) to observe at 330--800 MHz and the Karl. G. Jansky Very Large Array (VLA) to observe at 1--4 GHz. These observations - the ﬁrst of their kind - have led to the discovery of several non-ideal behaviours of ECME, which reveal the limitations of the current understanding of this phenomenon in hot magnetic stars.
%These observations, that are first of its kind, led to the discovery of several non-ideal behaviour of ECME, which reveal the limitations of the current understanding of the phenomenon in these hot magnetic stars.
%These observations led to the first detection of the magneto-ionic mode transition of ECME from hot magnetic stars, and also a few other surprising facts. The VLA observations also enabled us to measure the upper cut-off frequency of ECME for HD\,133880 which has not been accomplished before for any other hot magnetic star producing ECME.

This paper is structured as follows: in the next section (\S\ref{sec:obs}), we describe our observations and data analysis. This is followed by results (\S\ref{sec:results}) and discussions (\S\ref{sec:disc}). We end with conclusions in section (\S\ref{sec:conclusion}).

\section{Observations and data analysis}\label{sec:obs}
HD\,133880 was previously observed with the GMRT at 610 MHz with 33.3 MHz bandwidth \citep{das2018}. Although the idea was to obtain data for one complete rotation cycle, it could not be accomplished due to technical issues. The rotational phases that could not be observed happened to correspond to one of the magnetic nulls, where we expect to observe ECME. In order to fulfil our original goal, and to take advantage of the uGMRT, we reobserved the star in uGMRT band 4 (500-900 MHz) near its magnetic nulls. In addition, we also observed the star in uGMRT band 3 (300--500 MHz) to investigate the presence of ECME at even lower frequencies.

Around the same time, we took observations with the VLA in L (1--2 GHz) and S (2--4 GHz) bands in subarray mode near the expected phases of arrival of ECME (i.e. near the magnetic nulls). Table \ref{tab:obs} summarizes these new observations. The standard calibrator 3C286 was used for calibrating the absolute flux density scale and also for bandpass calibration in all the observations. The phase calibrators used for the uGMRT observations are J1517-243 and J1626-298. J1522-2730 was used as phase calibrator in the VLA observations.

\begin{table*}
{\scriptsize
\caption{Log of observations of HD\,133880 showing the dates and durations of observations at different wavebands and the effective observing frequency ranges (Eff. band) for each band on different days. The hours here refer to total observation hour including the overheads. Also shown are the rotational phase ranges ($\phi_\mathrm{rot}$) covered on different days. The rotational phases are calculated using Eq.\ref{eq:ephemeris}.\label{tab:obs}}
\begin{tabular}{cccc|cccc||cccc}
\hline
\multicolumn{8}{c}{uGMRT}       & \multicolumn{4}{c}{VLA}\\
\multicolumn{4}{c}{band 3} & \multicolumn{4}{c||}{band 4} &\multicolumn{4}{c}{L+S}\\
%\cline{2-12}
\hline
Date & Hours & Eff. band & $\Delta \phi_\mathrm{rot}$ & Date & Hours & Eff. band &$\Delta \phi_\mathrm{rot}$ & Date & Hours & Eff. band & $\Delta \phi_\mathrm{rot}$\\
& & MHz & &  & & MHz & &  & & GHz &\\
\hline
2019-03-17 & 6 & {\color{black}334--360}, & 0.63--0.88 & 2018-08-20 & 5 & 570--804 &0.13--0.32 & 2019-09-07 & 4 & 1.040--1.104, 1.360--1.488 & 0.13--0.30\\
 &  & {\color{black}380--461} & & 2019-08-02 & 5 &  570--804&0.61--0.81 & & & 1.680--2.000, 2.051--3.947 &\\
2019-05-17 & 6 & 334--461 & 0.10--0.35 &  &  &   & & 2019-11-01 & 4 & 1.040--1.104, 1.360--1.488 & 0.63--0.80\\
& & & & &  & & &  & & 1.680--2.000, 2.051--3.947 &\\
\hline
\end{tabular}
}
\end{table*}

The uGMRT data were analysed using the method described by \citet{das2019b}. For the VLA observations, the calibrated data were provided from the observatory itself via a pipeline processing. These data have 16 spectral windows in each band. The L band data were self calibrated without masking the target using the \textsc{casa} task `tclean'. No self-calibration was done for the S band data since there are not many sources at those frequencies and an image was made without masking the target. The sources other than the target were then subtracted using the task `uvsub' and the residual data were imaged with 2 minute time resolution for each spectral window.

To obtain the lightcurves, we have phased the data using the following ephemeris \citep{das2018}:
\begin{align}
\mathrm{HJD}&=2445472.0+0.877483\cdot E \label{eq:ephemeris}
\end{align}

\begin{figure}
\centering
\includegraphics[width=0.49\textwidth]{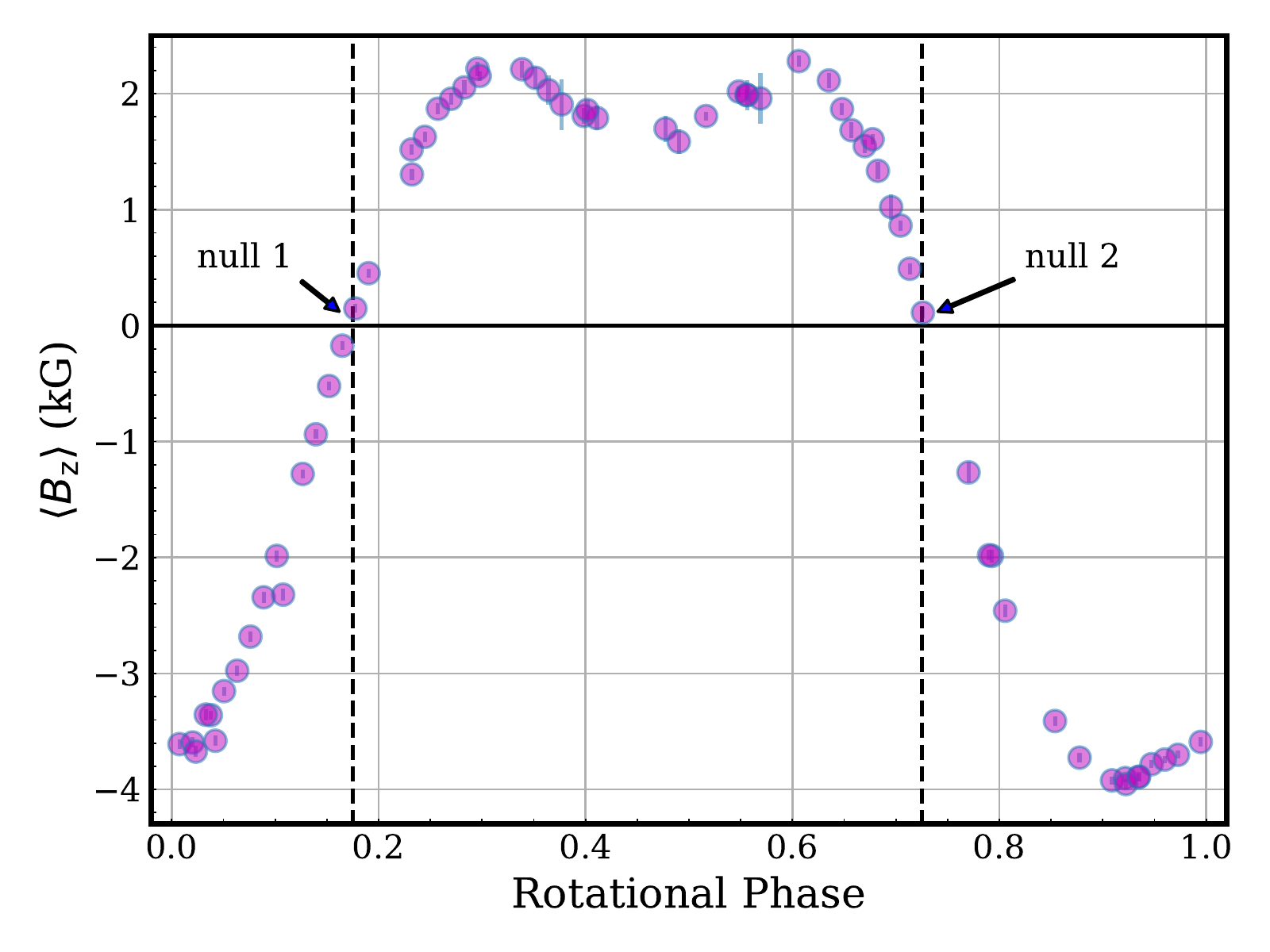}
\caption{The \bz~variation for this star as obtained by \citet{kochukhov2017}. The first null, where \bz~changes from negative to positive, lies at a rotational phase of $\approx0.175$. We will call it null 1. The second null, where \bz~changes from positive to negative, lies at $\approx0.725$ cycle. We will call it null 2.\label{fig:hd133880_bz}}
\end{figure}

Before we present our findings, we would like to highlight the fact that the convention for defining right and left circular polarizations used by \citet{das2018} corresponds to the one used by the GMRT observatory. It was recently discovered that this convention is opposite to the IEEE convention in band 4 (550--900 MHz) \citep{das2020}. By performing a similar experiment for uGMRT band 3, we arrived at the same conclusion. Thus, according to the IEEE convention, \citet{das2018} observed a nearly $100\%$ left circularly polarized pulse near the magnetic null where \bz~changes from positive to negative (null 2 in Figure \ref{fig:hd133880_bz}). 

In this paper, we use the IEEE convention for labelling right and left circular polarization. 
%Also hereafter, all mentioning about RCP and LCP are according to the IEEE convention unless stated otherwise.

\section{Results}\label{sec:results}

\begin{figure}
\centering
\includegraphics[width=0.49\textwidth]{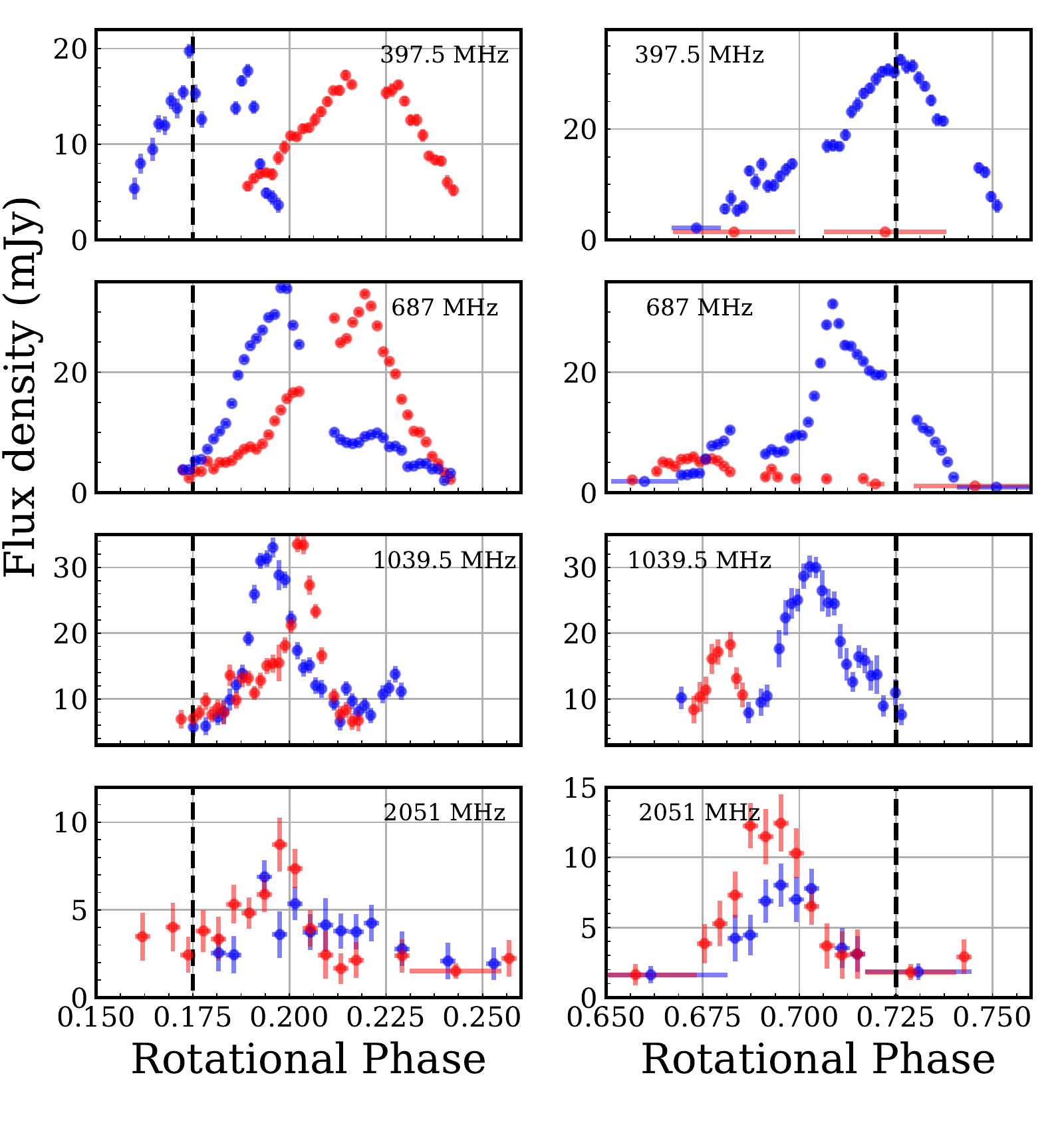}
\caption{The lightcurves of HD\,133880 in band 3, band 4, L and S bands near the two magnetic nulls. For L (1--2 GHz) and S (2--4 GHz) bands, we have shown the lightcurves for only one of the spectral windows (bottom two panels). The left panels are for null 1 and right panels are for null 2. Blue and red correspond to LCP and RCP respectively. {\color{black}The vertical dashed lines represent the magnetic null phases.}\label{fig:hd133880_lightcurves_combined}}
\end{figure}

As mentioned already, our observations cover only a small range of rotational phase near each magnetic null of the star. The \bz~curve for the star is shown in Figure \ref{fig:hd133880_lightcurves_combined}. For our convenience, we name the magnetic null where \bz~changes from negative to positive as null 1, and the one where \bz~changes from positive to negative as null 2. The lightcurves of HD\,133880, one pair (consisting of LCP and RCP) for each of the four wavebands, are shown in Figure \ref{fig:hd133880_lightcurves_combined}. The left panels show the lightcurves near null 1 and the right panels show those near null 2.
These observations reveal many new aspects of ECME from the star, each of which is described in the following subsections. 

\subsection{Confirming the presence of the RCP component of ECME from HD\,133880}\label{subsec:ecme_lcp}
\citet{das2018} confirmed the presence of ECME in HD\,133880 based on the presence of an intense, $\approx100\%$ LCP (according to the IEEE convention) pulse near null 2. In our new observations, we, for the first time, have observed RCP along with LCP. These results are illustrated in Figure \ref{fig:hd133880_lightcurves_combined}. However, the results differ greatly between the two nulls, and vary with frequency. Near null 1, the LCP and RCP pulses are of comparable amplitudes in all wavebands. But near null 2, RCP is not detected in band 3 (top right panel of Figure \ref{fig:hd133880_lightcurves_combined}); it is significantly weaker than LCP in band 4 (second right panel of Figure \ref{fig:hd133880_lightcurves_combined}); slightly weaker than LCP in the first spectral window of L band (1039.5 MHz), and comparable to LCP in the first spectral window of S band (2051 MHz). 

The non-detection of RCP pulse near null 2 in band 3 may be either due to the absence of any detectable enhancement in RCP, or due to the fact that we did not cover an adequate rotational phase range and the RCP pulse was therefore missed in our observations. The latter is a plausible explanation because we find that the separation between RCP and LCP pulses near null 2 is larger than that near null 1 at the same frequency (Figure \ref{fig:hd133880_lightcurves_combined}). Since in band 3, we see LCP before RCP near null 1, we expect to see RCP before LCP near null 2 \citep[e.g.][]{leto2016}. As the two null 2 pulses are relatively widely separated, the RCP pulse may lie outside of the observed rotational phase window.

\begin{figure}
\centering
\includegraphics[width=0.49\textwidth]{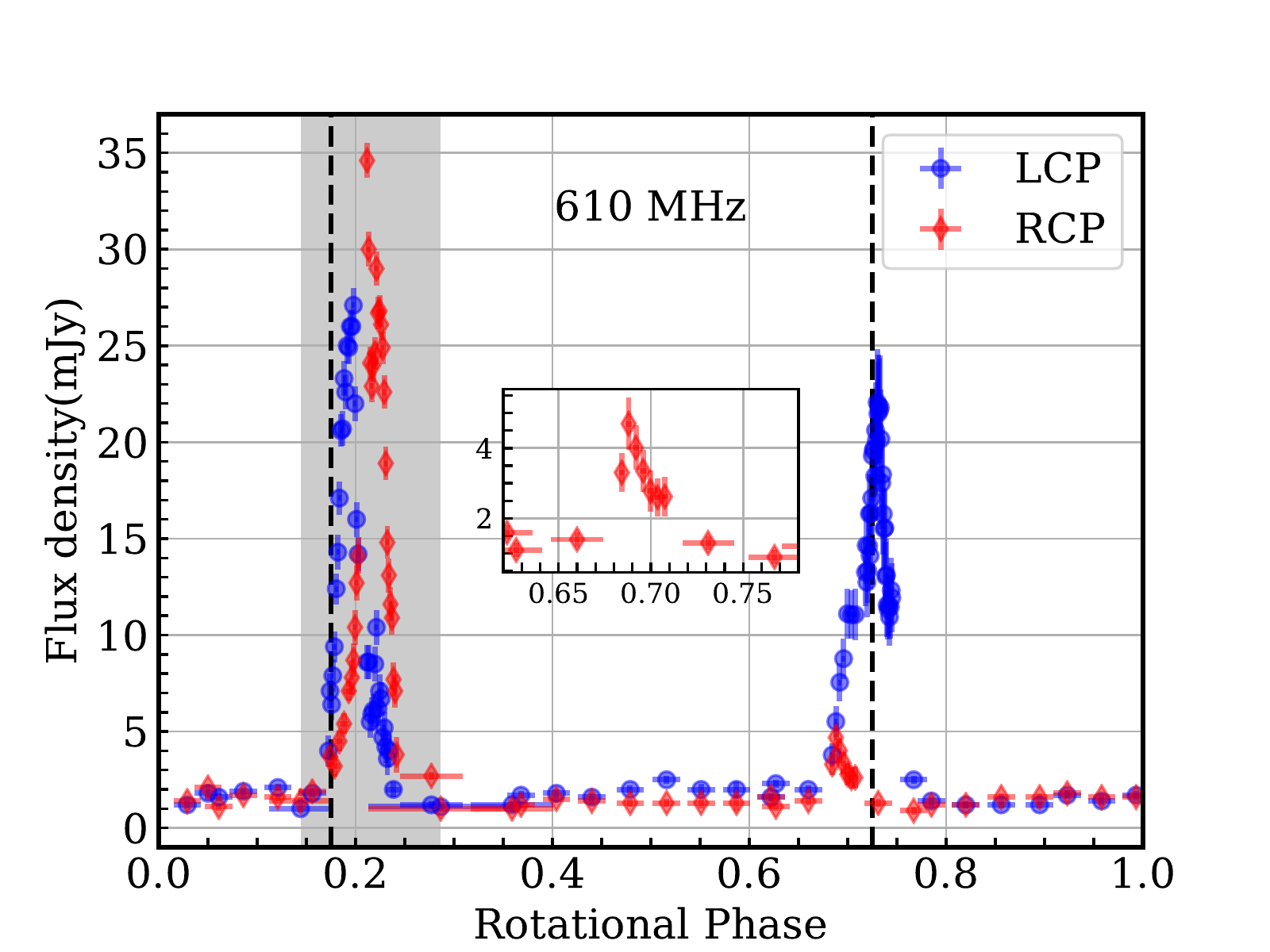}
\caption{The LCP (blue) and RCP (red) lightcurves of HD\,133880 for one comple rotation cycle at 610 MHz. Note that RCP and LCP are defined in accordance with the IEEE convention. The data within the region marked in grey are obtained with the uGMRT in band 4, whereas the remaining data were obtained with the GMRT with 33.33 MHz bandwidth, and reported by \citet{das2018}. The dashed line correspond to the magnetic nulls. The weak enhancement in RCP observed in the data of \citet{das2018} near null 2 is shown in the inset.\label{fig:hd133880_610mhz_full_lightcurve}}
\end{figure}

\begin{figure}
\centering
\includegraphics[width=0.49\textwidth]{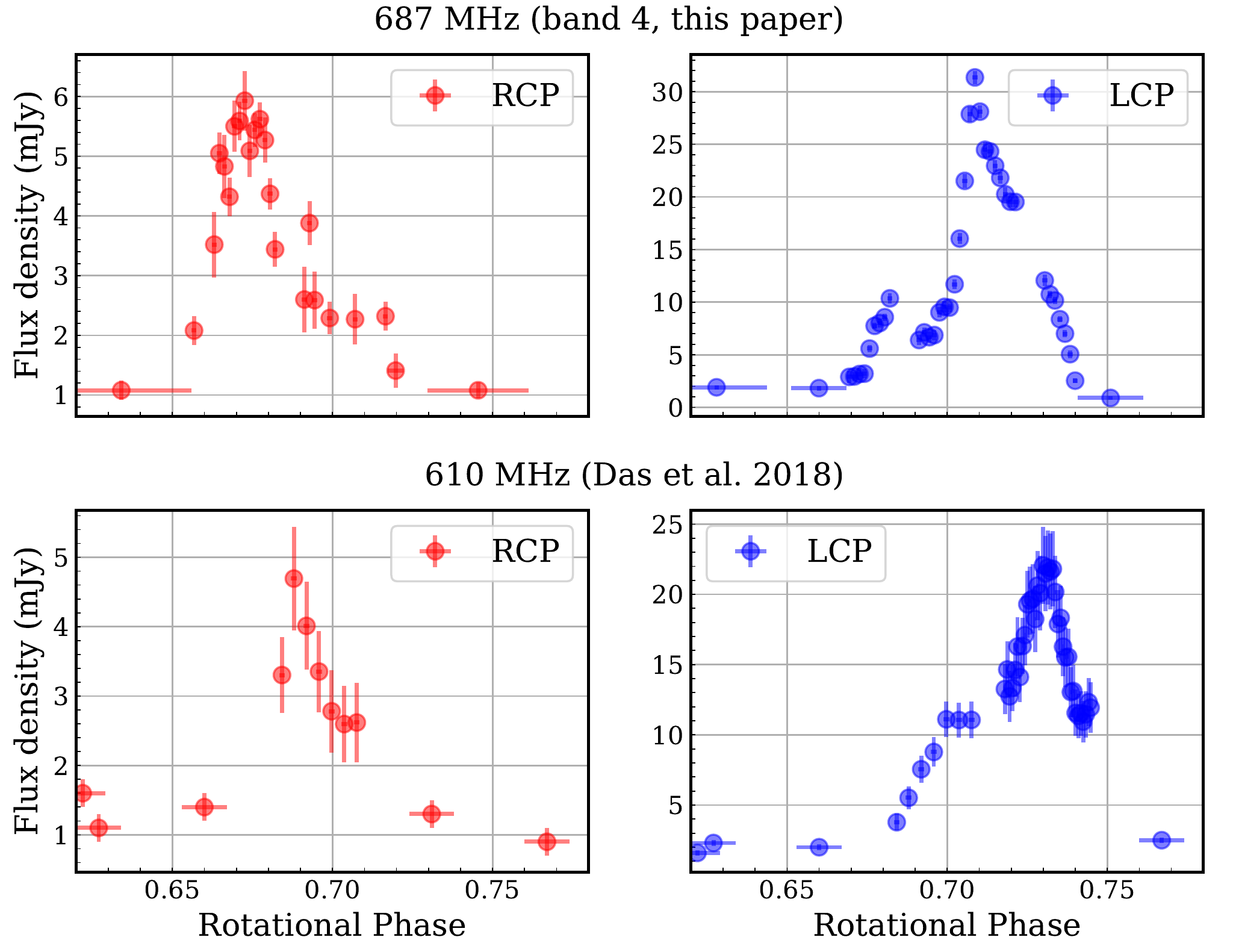}
\caption{\textit{Upper panels:} The lightcurves of HD\,133880 over the frequency range of band 4 (687 MHz, 234 MHz bandwidth) near null 2 where \bz~changes from positive to negative. These data were acquired in August 2019 with the uGMRT. The left panel shows the RCP lightcurve and the right panel shows the LCP lightcurve. \textit{Lower panels:} Same as the upper panels, but for the data reported by \citet{das2018}. These data were taken in February 2016 with the GMRT at 610 MHz (33.33 MHz bandwidth). Note that the range of the Y axes are different for the left and right panels.\label{fig:hd133880_band4_610MHz_null2}}
\end{figure}

One interesting outcome of these observations is the possible presence of RCP enhancement in the observations of \citet{das2018}.
We have reproduced the upper panel of their Figure 1 in Figure \ref{fig:hd133880_610mhz_full_lightcurve} with the circular polarizations now labelled according to the IEEE convention, and supplemented it with our new data near null 1.
We find that there was actually a very weak enhancement in RCP just before the LCP pulse observed near null 2.
It was not associated with ECME by \citet{das2018} mainly because of its weak strength. However in light of the new data, the weak RCP enhancement observed near null 2 acquires a new significance since it appeared before the LCP pulse which is expected for ECME near null 2. This idea is further strengthened by the result of our re-observation near null 2 in band 4 on 2019 August 2. In Figure \ref{fig:hd133880_band4_610MHz_null2}, we show the lightcurves near null 2 obtained from our re-observation in band 4 (upper panels) and those reported by \citet{das2018} (lower panels). The RCP lightcurves are shown in the left panels and those in LCP are shown in the right panels. We can clearly see that the rotational phase range over which the enhancement in RCP is seen in the new data is roughly the same as that where the weak RCP enhancement in the data of \citet{das2018} lies. In both cases, the RCP pulse is significantly weaker than the LCP pulse. 

\subsection{Possible evolution of the rotation period of HD\,133880}\label{subsec:_period_evolution}
\begin{figure}
\centering
\includegraphics[width=0.49\textwidth]{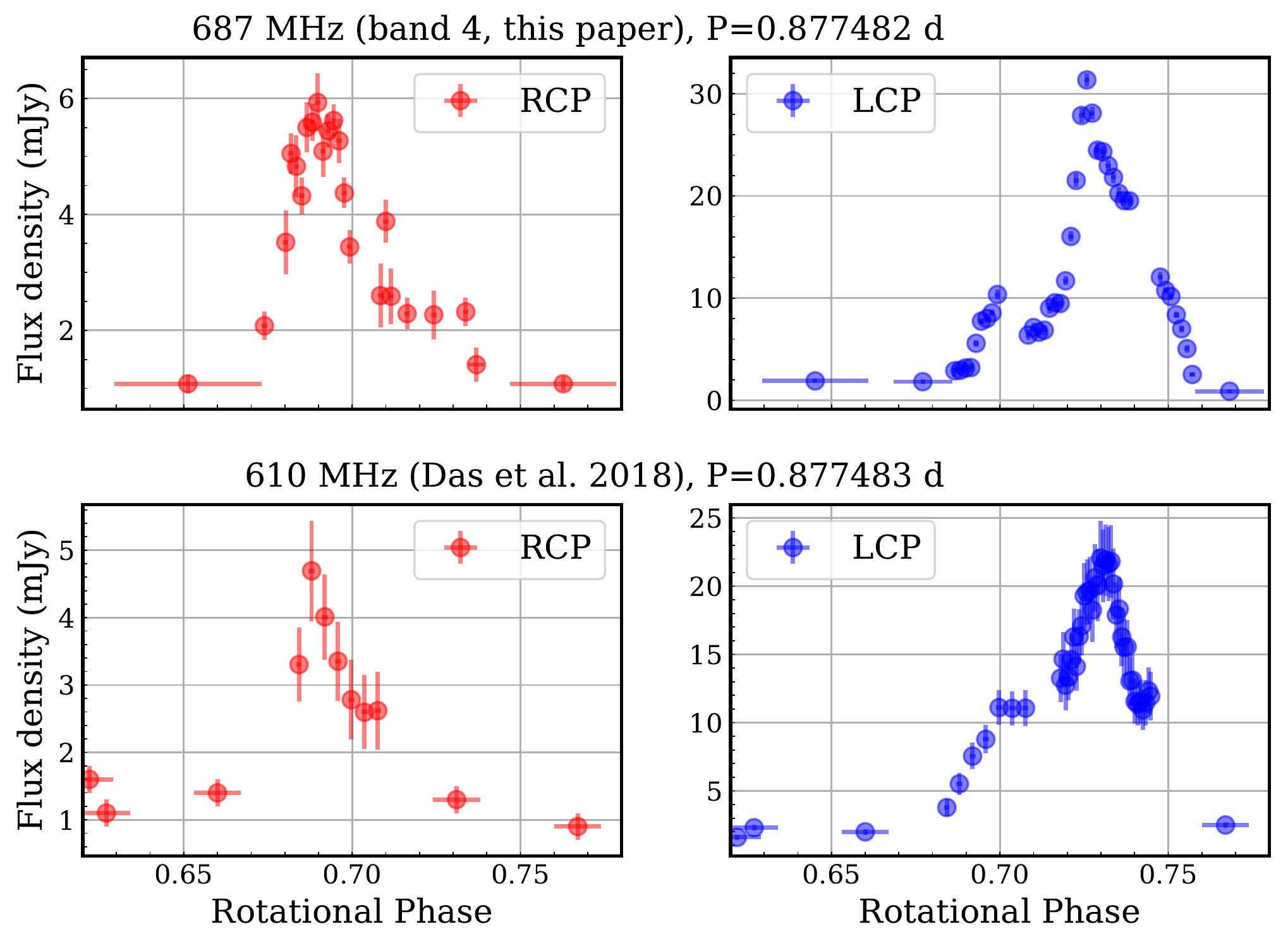}
\caption{Same as Figure \ref{fig:hd133880_band4_610MHz_null2}, but the new data (presented in this paper) are phased with a rotation period 0.877482 days (upper panels) and those of \citet{das2018} are phased with a rotation period of 0.877483 days (lower panels). Note that the range of the Y axes are different for the left and right panels.\label{fig:hd133880_period_change}}
\end{figure}

A byproduct that has resulted from our observation of the star in band 4 near null 2 is that we can compare the rotational phases of arrival of the ECME pulses observed in our data (taken on 2019 August 2) with those obtained by \citet{das2018} which were acquired on 2016 February 8. Upon doing so, we find from Figure \ref{fig:hd133880_band4_610MHz_null2} that the enhancements in the new data are shifted by $\approx 0.02$ cycle relative to the enhancements reported by \citet{das2018}. This discrepancy can be resolved if we adopt a rotation period of 0.877482 days instead of 0.877483 days \citep[suggested by][]{kochukhov2017} for the new data, with the old data phased using the latter period (Figure \ref{fig:hd133880_period_change}). This could suggest an evolution of the stellar rotation period by $\approx-0.09$ s between the years 2016 and 2019. 
Note that rotation period evolution has been suggested for other hot magnetic stars producing ECME \cite[e.g. CU\,Vir and HD\,142990,][]{trigilio2008,mikulasek2011,shultz2019_0}. 
{\color{black}Although spin-down of magnetic stars is expected to happen due to magnetic braking \citep[e.g.][]{ud-doula2009}, spin-up is a surprising phenomenon, which is, though have been observed for a number of stars, is a puzzle not fully solved. Possible mechanisms for spin-up of a magnetic star includes stellar torsional oscillations produced by interaction of the internal magnetic field and differential rotation \citep{krticka2017}, and ongoing core-contraction of the star \citep{shultz2019_0}. }

\subsection{The magneto-ionic mode of ECME }\label{subsec:ecme_mode}
The significance of the discovery of the RCP component of ECME lies in the fact that we can now infer the magneto-ionic mode of ECME directly from the observations. The magneto-ionic mode can be determined from the sequence of arrival of LCP and RCP pulses near the two nulls \citep{leto2019}. The difference between the lightcurves in X-mode and O-mode ECME is illustrated in Figure 1 of \citet{das2019a}. Using this strategy, we find from our Figure \ref{fig:hd133880_lightcurves_combined} that the magneto-ionic mode is X-mode over the entire frequency range of our observations. From this conclusion, we estimate that the plasma density at the height at which ECME at 400 MHz is produced is $<2\times10^8\,\,\mathrm{cm^{-3}}$ for emission at the fundamental harmonic, and 
$<0.4\times10^8\,\,\mathrm{cm^{-3}}$ for emission at the second harmonic \citep{sharma1984,melrose1984,leto2019}.

{\color{black}\subsection{The spectra of ECME}\label{subsec:ecme_cutoff}}
\begin{figure*}
\centering
\includegraphics[width=0.95\textwidth]{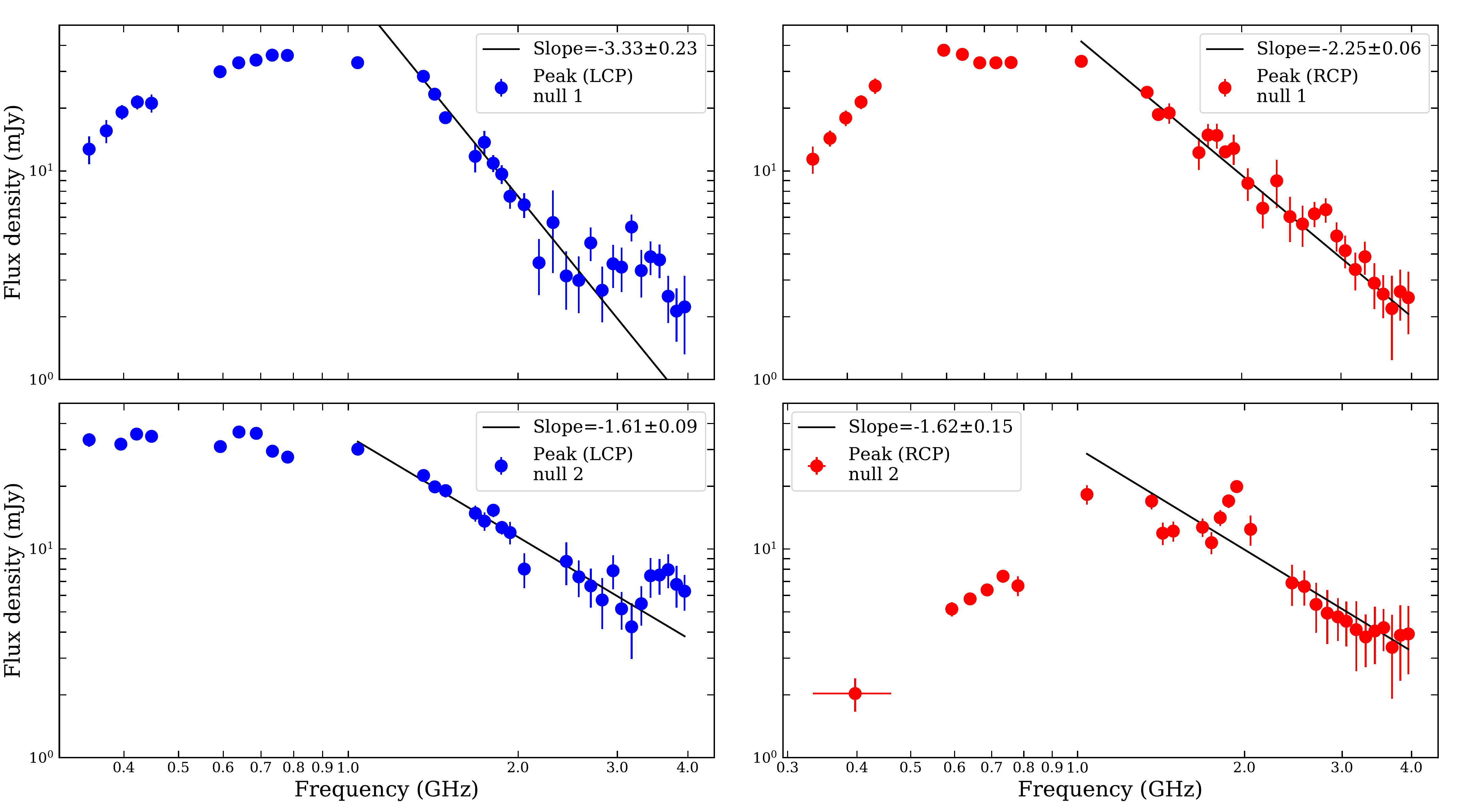}
\caption{{\color{black}The spectra of ECME for the LCP (blue markers) and RCP (red markers) pulses near the null 1 (upper panel) and null 2 (lower panel). The flux density plotted here corresponds to the maximum observed flux density at that frequency.}\label{fig:hd133880_spectra}}
\end{figure*}

{\color{black}In Figure \ref{fig:hd133880_spectra} we show the peak ECME flux density against frequency for the RCP and LCP pulses observed near the two magnetic nulls. To obtain a better idea about the variation of flux density with frequency over the uGMRT bands 3 and 4, we divided the two bands into smaller sub bins. Band 4 was divided into five sub-bins, each of width $\approx 47\,\mathrm{MHz}$. Band 3 was also divided into five sub-bins for the data taken on 2019 May 17 (null 1, Table \ref{tab:obs}), each of width $\approx 25\,\mathrm{MHz}$. For the band 3 data taken on 2019 March 17 (null 2, Table \ref{tab:obs}), we could get only four sub-bins due to a larger flagging of data. Also, the sub-division was done only for the LCP pulse, since we did not observe any enhancement in RCP. 

From Figure \ref{fig:hd133880_spectra} we find that all the spectra show a power-law like tail, the onset of which occurs at $\gtrsim 1\,\mathrm{GHz}$. By fitting a power-law to the data between 1--4 GHz, we find that the spectral indices are different for the RCP and LCP pulses observed near null 1, but similar for the pulses observed near null 2. We also find that the spectra for the pulses near null 1 are much steeper than those for the pulses observed near null 2 above 1 GHz.

The spectra for the pulses near null 1 also show a rising part between 300--600 MHz and then a turn-over between 0.6--1.4 GHz, followed by the power-law decline. The spectrum for the RCP pulse observed near null 2 also has a rising part, but located between 300--1000 MHz, with the turn-over happening between 1--2 GHz. On the other hand, the spectrum for the LCP pulse near null 2 remains essentially flat between 0.3--1 GHz.

These results suffer from one crucial limitation, which is that the data obtained in different frequency bands are not simultaneous. ECME flux density observed from CU\,Vir is known to vary with time \citep[e.g.][]{trigilio2011}, and the same may be true for HD\,133880. One observation that suggests that this may not be true is Figure \ref{fig:hd133880_band4_610MHz_null2}, where we show the 610 MHz pulses near null 2 observed on two epochs separated by three years. The flux densities, though not identical, are not dramatically different between the two epochs. This however needs to be confirmed by further re-observations in band 4 and in the VLA L and S bands. Hence, at this point, though we are confident about the intraband flux density variation (in this context, the VLA L and S bands constitute a single band since the data were taken in sub-array mode), we cannot be sure of the interband spectral shapes. The spectral indices, obtained using only the VLA data, are however robust against this limitation.
}

\subsection{Cut-off frequency of ECME}\label{subsec:ecme_cutoff}
\begin{figure*}
\centering
\includegraphics[width=0.85\textwidth]{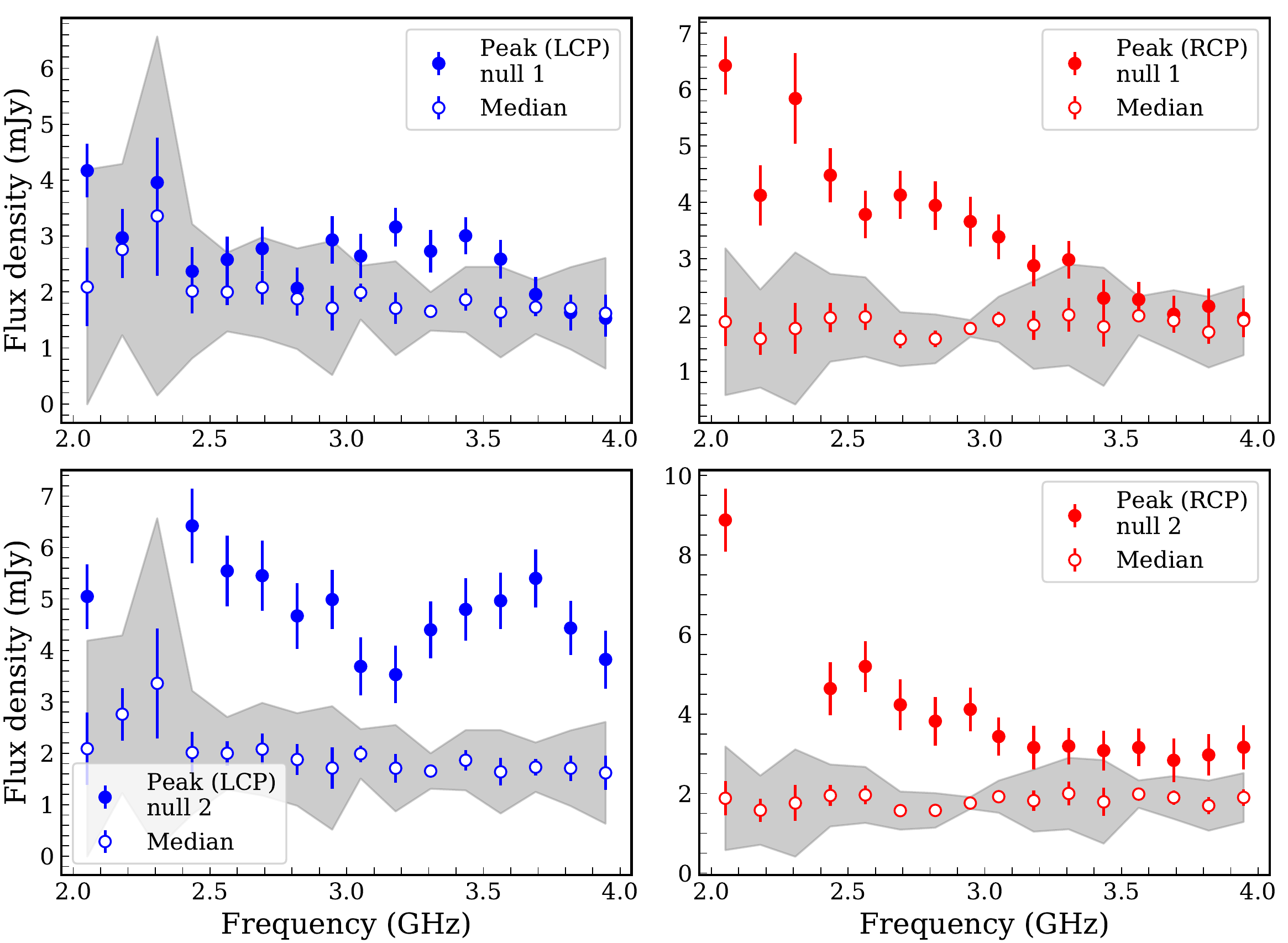}
\caption{The variation of peak flux density (filled markers) with frequency in RCP (red markers) and LCP (blue markers) near the two nulls. Also shown is the variation of median basal flux density (unfilled markers) with frequency. For definitions of peak and median basal flux density, refer to \S\ref{subsec:ecme_cutoff}. On 2019 November 1 (observation near null 2, bottom panels), we lost two spectral windows centred at 2.2 and 2.3 GHz due to severe radio frequency interference (RFI), which is why there are not any data for peak flux density at these frequencies near null 2.\label{fig:hd133880_S_peak_flux_vs_freq}}
\end{figure*}

It is predicted that a cut-off frequency should exist, above which the ECME phenomenon should cease. Although some constraints on the cut-off frequency have been proposed \citep[e.g.][]{leto2019}, no direct measurement has been obtained for any of the six hot magnetic stars known to exhibit ECME.
For HD\,133880, it is known that ECME is absent at 5 GHz \citep{lim1996} which imposes an upper limit to the cut-off frequency of ECME. Through our new observations over a continuous frequency range of 2--4 GHz (S band of the VLA), we have been able to trace for the first time the way the phenomenon dies out with increasing frequency. To do that, we construct a plot of peak and median basal flux densities against frequency. 
{\color{black}The peak flux density is obtained by first smoothing the lightcurves by averaging over 5 data points (or 10 minutes) and then finding the maximum.} This is done to take into account that the ECME pulses have certain widths, and a single point of high flux density is unlikely to be of ECME origin.
The median basal flux density is obtained by taking the median of the flux densities over a rotational phase range devoid of any enhancement, which we choose to be 0.22--0.30 cycle. The error bars in the median values are the median absolute deviations (MAD). The results are shown in Figure \ref{fig:hd133880_S_peak_flux_vs_freq}. Blue and red markers correspond to LCP and RCP respectively. The grey regions around the medians correspond to $3\times\mathrm{MAD}$. We will define the cut-off frequency as the lowest frequency where the peak flux density overlaps with the grey region within the errorbars. Thus for LCP and RCP near null 1, the cut-off frequencies are $\approx2.0$ GHz and $\approx3.2$ GHz respectively. That for RCP near null 2 is also $\approx 3.2$ GHz. However for LCP near null 2, no cut-off was detected according to this definition. In that case the cut-off must lie between 4--5 GHz. 
%The fact that the cut-off frequencies are different for different pulses is surprising and unexpected in the ideal case. We will discuss this point briefly in \S\ref{sec:disc}.
The different cut-off frequencies observed for different pulses could be an artefact of our definition since the MAD for different polarizations are different, and also varies with frequencies. However by comparing pulses at the same polarizations, we infer that whereas the RCP pulses near both nulls behave similarly with frequencies, the same is not true for the LCP pulses. 

{\color{black}Similar to the upper cut-off frequency, there should also be a lower cut-off frequency below which ECME will be absent. From the top panels and the bottom left panels of Figure \ref{fig:hd133880_spectra}, it appears that the lower cut-off frequency for the pulses near null 1, and for the LCP pulse near null 2, is $<350$ MHz. However for the RCP pulse near null 2, the data suggest that the low frequency cut-off is around 400 MHz. This can however only be confirmed by re-observing over a broader rotational phase range for the reason mentioned in \S\ref{subsec:ecme_lcp}.}

%We also note that our observations also achieve the lowest frequency detection of ECME from HD\,133880, which is $\approx 400$ MHz. This frequency serves as an upper limit to the lower cut-off frequency of ECME from this star.

\section{Discussion}\label{sec:disc}
In this paper we present wideband observations of the magnetic B star HD\,133880 over the frequency range of {\color{black}300--800} MHz and 1--4 GHz {\color{black}near the magnetic null phases}. 
Our observations confirm that the {\color{black}lower cut-off frequency of ECME for this star is $<350$ MHz for the pulses near null 1, and the LCP pulse near null 2; but it may be around 400 MHz for the RCP pulse near null 2.} The upper cut-off lies at around 3 GHz for both RCP and LCP pulses near null 1 and also the RCP pulse near null 2, but for the LCP pulse near null 2, the cut-off seems to be higher, and to lie between 4--5 GHz according to our definition of the cut-off described in \S\ref{subsec:ecme_cutoff}. We also find that both LCP and RCP components of ECME are present over our observing frequency range. {\color{black}For the pulses coming from the south magnetic pole (the LCP pulses), The observed frequency range of ECME, i.e. $\approx 0.3-3$ GHz, corresponds to 1.2--3.7 stellar radii and 1.7--4.9 stellar radii above the surface of the star for ECME at the fundamental and second harmonic respectively. In case of the RCP pulses coming from the north magnetic pole, the corresponding heights are 0.6-2.5 stellar radii and 1.0-3.4 stellar radii above the surface of the star for ECME at the fundamental and second harmonic respectively. The heights for the RCP and LCP pulses are different because of the distorted topology of the stellar magnetic field \citep{kochukhov2017}.}

In addition to these findings, there are several peculiar characteristics which came to light through these observations. 
One such peculiarity is the difference in strength of the RCP and LCP pulses near null 2. At lower frequencies, the RCP pulse is significantly weaker than the LCP pulse. This difference gradually decreases at higher frequencies and eventually vanishes around 1.5 GHz. Since near null 2 \bz~changes from positive to negative, the RCP pulse, that arrives first, corresponds to the North pole regions. Note that the second pulse observed near null 1 is also from the North pole regions \citep[see \S2 of][]{das2019a}, but seen at a different stellar orientation. Thus instability at the emission site, which is usually thought to be the reason behind pulse-strength variability \citep[e.g.][]{trigilio2011} cannot explain why the pulse coming from the North pole is \textbf{always} weak when viewed at a certain orientation of the stellar magnetosphere, since such instability, if there is any, is expected to be random and should not depend on the stellar orientation. Therefore, the cause behind the persistently weaker pulse near null 2 appears to be external and unrelated to the emission site. Because this effect is dependent on the stellar orientation, it points towards a magnetosphere which is not azimuthally symmetric about the magnetic field axis. This is not particularly surprising given that the magnetic field of the star is already known to be a distorted, asymmetric dipole \citep{kochukhov2017}. Even an azimuthally symmetric dipolar magnetic field does not guarantee an azimuthally symmetric density distribution in the magnetosphere unless the rotation axis and the magnetic field axis are aligned \citep[the `Rigidly Rotating Magnetosphere' model,][]{townsend2005}, which is clearly not the case for HD\,133880 \citep[the angle between the rotation axis and magnetic field axis, or obliquity, is $78^\circ$,][]{kochukhov2017}. Thus, from the high obliquity alone, we can expect a highly azimuthally asymmetric (about the magnetic field axis) distribution of matter which can lead to different amounts of absorption/refraction suffered by the radiation at different rotational phases. {\color{black} This has been shown to be the case by \citet{das2020b} through simulation. They found that propagation effects in an azimuthally asymmetric magnetosphere can influence the pulses observable at different nulls and different frequencies by different amounts, and can also lead to a complete disappearance of a pulse from a particular magnetic polar region.}
Thus, the high obliquity combined with the non-ideal dipolar magnetic topology of the star can give rise to a highly asymmetric magnetosphere and the weak pulse near null 2 is possibly one of the signatures of that. 

Once we accept that the reality is far from the ideal azimuthally symmetric magnetosphere, the other peculiarities that we observed no longer remain surprises. For example, we observed different separations between the LCP and RCP pulses near the two nulls at the same frequency. The separation between the RCP and LCP components is believed to be caused by the refraction suffered in the dense plasma in the inner magnetosphere \citep{trigilio2011,leto2016}. If the plasma density is not azimuthally symmetric, the pulses corresponding to the two nulls will suffer different amounts of deviation for the two orientations and consequently the separations will also be different. 
{\color{black}We also find that the rotational phases of arrival of the pulses are not symmetric about the magnetic nulls (Figure \ref{fig:hd133880_lightcurves_combined}). This can also be explained from propagation effects in a `non-ideal' magnetosphere \citep{das2020b}.}
Similarly the different high frequency cut-offs for the different components of ECME can also be due to the azimuthal asymmetry, but only if the origin of this cut-off is imaginary refractive index due to high plasma density
\citep{leto2019}. In that case, it indicates presence of plasma with number density $>10^{11}\,\,\mathrm{cm^{-3}}$ in the stellar magnetosphere. We would however like to point out that our definition of the cut-off frequency has the limitation that it is not uniform for the two circular polarizations.

%Similarly the different high frequency cut-offs for the different components of ECME can also be due to the azimuthal asymmetry. Our observation that the LCP pulses, coming from the same magnetic polar regions, have different upper cut-off frequencies near the two magnetic nulls, implies a dependence of the phenomenon on the stellar orientation w.r.t. the observer. This suggests that the reason for the cur-off is not intrinsic to the emission site, but probably due to absorption of the radiation by dense plasma in the stellar magnetosphere. This is also proposed by \citet{leto2019}.

Even after assuming an azimuthally asymmetric density profile around the magnetosphere, we cannot explain the misalignment observed between pulses near null 2 at two epochs (Figure \ref{fig:hd133880_band4_610MHz_null2}) without considering an evolution of the rotation period. The slight difference in frequency (610 MHz and 687 MHz) of the pulses cannot explain the shift as the effect of this difference is to shift the LCP and RCP in opposite directions (e.g. Figure \ref{fig:hd133880_lightcurves_combined}).

{\color{black}One of the important results of this paper is the wideband spectra of ECME for the RCP and LCP pulses near the two magnetic nulls. We find that the spectra exhibit power-law tails, but the spectral indices are not the same for the oppositely circularly polarized pulses near null 1, and also for the pulses near the two nulls. The difference in spectral index between RCP and LCP pulses near null 1 is not surprising since the RCP and LCP pulses have different sites of emission. However, the fact that the spectral indices corresponding to the LCP (or RCP) pulses, observed near the two nulls, are different, is surprising since they come from the same emission site. In future, it will be interesting to investigate if the spectral shapes evolve with time.}

{\color{black} Another interesting aspect of ECME that is worth exploring in the future is to try to disentangle the effects of absorption and refraction on the observed ECME pulses. The first step in this direction will be to check the time evolution of the spectra, both in terms of shape and intensity. As demonstrated by \citet{das2020b}, the observed spectra are affected by both refraction and absorption, and hence the spectra alone cannot help us to disentangle the two effects. This problem might be possible to overcome by modelling the associated `lag's (differences in arrival time of pulses at different frequencies), which is believed to be entirely a result of refraction \citep{trigilio2011}. Once we get an idea about refraction, we can remove its contribution from the observed spectra. The next step is also not straight-forward since we do not know the spectrum of ECME at the time of emission. In that case, we may assume the spectrum of a pulse observed near one of the nulls as the original spectrum, and then determine if absorption is needed to produce the spectrum observed for the same pulse near the other magnetic null. For example, we can consider the spectrum for the RCP pulse near null 1 as the `true' spectrum, and then try to find out the importance of absorption to produce the spectrum for the RCP pulse near null 2.
}

To summarize, the wideband observations of HD\,133880 near its magnetic nulls {\color{black}unravel a number of previously unexplored aspects of ECME, and} reveal a significant deviation of the distribution of matter from the ideal azimuthally symmetric distribution in the stellar magnetosphere.

\section{Conclusion}\label{sec:conclusion}
The primary conclusions drawn from our wideband observations of HD\,133880 over the frequency range of 330--800 MHz and 1--4 GHz are the following:

\begin{enumerate}
\item The magneto-ionic mode of ECME over the frequency range of observation is extra-ordinary. From this inference, we estimate the plasma density at the site of origin of ECME at 400 MHz to be $<2\times10^8\,\,\mathrm{cm^{-3}}$ for emission at the fundamental harmonic, and $<0.4\times10^8\,\,\mathrm{cm^{-3}}$ for emission at the second harmonic. 

\item Both the LCP and RCP components of ECME are present near both magnetic nulls in band 4, L and S bands. In band 3, we did not observe the RCP pulse near null 2. {\color{black}This could be either due to a low frequency cut-off, or an inadequate rotational phase coverage.}

\item {\color{black}The shapes of the ECME spectra for the RCP and the LCP pulses observed near the two magnetic nulls are different. However, all four spectra exhibit power-law tails. The power-law is steeper for the pulses observed near null 1 than those obtained for the pulses near null 2.}

\item The distribution of magnetospheric plasma density deviates significantly from azimuthal symmetry about the magnetic field axis. This is consistent with the star's high obliquity and non-ideal dipolar magnetic field \citep{townsend2005,kochukhov2017}.

\item The upper cut-off frequency of ECME is different for the different components of ECME according to our definition (\S\ref{subsec:ecme_cutoff}). This could also be a signature of the azimuthally asymmetric matter distribution in the magnetosphere, provided the reason for the cut-off is imaginary refractive index due to high plasma density \citep{leto2019}.
If this is the case, there must be plasma with number density $>10^{11}\,\,\mathrm{cm^{-3}}$.

%For the pulses near null 1 and also for the RCP pulse near null 2, it lies between 2--4 GHz, whereas for the LCP pulse near null 2, the cut-off probably lies between 4--5 GHz. The dependence of cut-off frequency on stellar orientation suggests that the reason for cut-off is probably not intrinsic to the emission site. An imaginary refractive index due to high plasma density can explain this property \citep{leto2019}, given that the magnetosphere appears to have a highly asymmetric density distribution. In that case, there must exist dense plasma in the magnetosphere with number density $>10^{11}\,\,\mathrm{cm^{-3}}$.

\item The current ephemeris for the star cannot phase ECME from 2016 and 2019 consistently. Either the adopted rotation period is wrong, or it has evolved. If the rotation period has evolved, our data suggests that the rate of change of the rotation period is {\color{black}$\approx -0.025\,\mathrm{s/yr}$} assuming a uniform rate of change over the years 2016 and 2019.

\end{enumerate}

In the future we aim to study the link between the density profile of the inner magnetosphere and the differences in various properties of ECME near the two nulls. This way, ECME can become a powerful tool to constrain the density distribution in magnetospheres which are far from ideal.

\section*{Data availability}
The uGMRT data used in this article are available in \url{https://naps.ncra.tifr.res.in/goa/data/search} under proposal codes ddtC010, 35\_097 and 36\_034. The VLA data are available in \url{https://archive.nrao.edu/archive/advquery.jsp} under the project code 19A-130. The data were analyzed using \textsc{casa} \citep{mcmullin2007}.

\section*{Acknowledgements}
We thank C. Trigilio for highly insightful and thought-provoking suggestions and comments, that helped us to improve this manuscript significantly. We acknowledge support of the Department of Atomic Energy, Government of India, under project no. 12-R\&D-TFR-5.02-0700. PC acknowledges support from the Department of Science and Technology via 
SwarnaJayanti Fellowship awards (DST/SJF/PSA-01/2014-15). GAW acknowledges Discovery Grant support from the Natural Sciences and Engineering Research Council (NSERC) of Canada. We thank the staff of the GMRT and the National Radio Astronomy Observatory (NRAO) that made our observations possible. 
The GMRT is run by the National Centre for Radio Astrophysics of the Tata Institute 
of Fundamental Research. The National Radio Astronomy Observatory is a facility of the National Science Foundation operated under cooperative agreement by Associated Universities, Inc. This research has made use of NASA's Astrophysics Data System.

%%%%%%%%%%%%%%%%%%%%%%%%%%%%%%%%%%%%%%%%%%%%%%%%%%

%%%%%%%%%%%%%%%%%%%% REFERENCES %%%%%%%%%%%%%%%%%%

% The best way to enter references is to use BibTeX:

\bibliographystyle{mnras}
\bibliography{das} % if your bibtex file is called example.bib

%%%%%%%%%%%%%%%%%%%%%%%%%%%%%%%%%%%%%%%%%%%%%%%%%%

%%%%%%%%%%%%%%%%% APPENDICES %%%%%%%%%%%%%%%%%%%%%

%%%%%%%%%%%%%%%%%%%%%%%%%%%%%%%%%%%%%%%%%%%%%%%%%%

% Don't change these lines
\bsp	% typesetting comment

\label{lastpage}
\end{document}